\documentclass[aps,prl,twocolumn,superscriptaddress,showpacs,showkeys,amsmath,amssymb,floatfix]{revtex4}%preprint <-> twocolumn
\usepackage{booktabs,graphicx,mathrsfs,verbatim}
\usepackage[colorlinks=true,citecolor=blue,filecolor=blue,linkcolor=blue,urlcolor=blue,pdftex]{hyperref}
\usepackage[usenames,dvipsnames]{color}
\usepackage{ulem}

\newcommand{\1}[1]{\, \mathrm{#1}} % unit(y ;-)
\newcommand{\n}[1]{\mathrm{#1}} % normal (roman) text in math mode

\newcommand{\percent}{\%}

\newcommand{\earth}{\oplus}

\newcommand{\assergi}{\affiliation{INFN Laboratori Nazionali del Gran Sasso, Assergi, 67100, Italy}}
\newcommand{\bologna}{\affiliation{University of Bologna and INFN-Bologna, Bologna, Italy}}
\newcommand{\columbia}{\affiliation{Physics Department, Columbia University, New York, NY 10027, USA}}
\newcommand{\coimbra}{\affiliation{Department of Physics, University of Coimbra, R. Larga, 3004-516, Coimbra, Portugal}}
\newcommand{\heidelberg}{\affiliation{Max-Planck-Institut f\"ur Kernphysik, Saupfercheckweg 1, 69117 Heidelberg, Germany}}
\newcommand{\houston}{\affiliation{Department of Physics and Astronomy, Rice University, Houston, TX 77005 - 1892, USA}}

\newcommand{\losangeles}{\affiliation{Physics \& Astronomy Department, University of California, Los Angeles, USA}}
\newcommand{\mainz}{\affiliation{Institut f\"ur Physik, Johannes Gutenberg Universit\"at Mainz, 55099 Mainz, Germany}}
\newcommand{\munster}{\affiliation{Institut f\"ur Kernphysik, Westf\"alische Wilhelms-Universit\"at M\"unster, 48149 M\"unster, Germany}}
\newcommand{\shanghai}{\affiliation{Department of Physics, Shanghai Jiao Tong University, Shanghai, 200240, China}}
\newcommand{\subatech}{\affiliation{SUBATECH, Ecole des Mines de Nantes, CNRS/In2p3, Universit\'e de Nantes, 44307 Nantes, France}}
\newcommand{\weizmann}{\affiliation{Department of Particle Physics and Astrophysics, Weizmann Institute of Science, 76100 Rehovot, Israel}}
\newcommand{\zurich}{\affiliation{Physics Institute, University of Z\"{u}rich, Winterthurerstr. 190, CH-8057, Switzerland}}

\begin{document}

\title{Implications on Inelastic Dark Matter from 100 Live Days of XENON100 Data}

\author{E.~Aprile}\columbia
\author{K.~Arisaka}\losangeles
\author{F.~Arneodo}\assergi
\author{A.~Askin}\zurich
\author{L.~Baudis}\zurich
\author{A.~Behrens}\zurich
\author{K.~Bokeloh}\munster
\author{E.~Brown}\munster
\author{T.~Bruch}\zurich
\author{G.~Bruno}\assergi
\author{J.~M.~R.~Cardoso}\coimbra
\author{W.-T.~Chen}\subatech
\author{B.~Choi}\columbia
\author{D.~Cline}\losangeles
\author{E.~Duchovni}\weizmann
\author{S.~Fattori}\mainz
\author{A.~D.~Ferella}\zurich
\author{F.~Gao}\shanghai
\author{K.-L.~Giboni}\columbia
\author{E.~Gross}\weizmann
\author{A.~Kish}\zurich
\author{C.~W.~Lam}\losangeles
\author{J.~Lamblin}\subatech
\author{R.~F.~Lang}\columbia
\author{C.~Levy}\munster
\author{K.~E.~Lim}\columbia
\author{Q.~Lin}\shanghai
\author{S.~Lindemann}\heidelberg
\author{M.~Lindner}\heidelberg
\author{J.~A.~M.~Lopes}\coimbra
\author{K.~Lung}\losangeles
\author{T.~Marrod\'an~Undagoitia}\zurich
\author{Y.~Mei}\houston\mainz
\author{A.~J.~Melgarejo~Fernandez}\email{ajmelgarejo@astro.columbia.edu}\columbia
\author{K.~Ni}\shanghai
\author{U.~Oberlack}\houston\mainz
\author{S.~E.~A.~Orrigo}\coimbra
\author{E.~Pantic}\losangeles
\author{R.~Persiani}\bologna
\author{G.~Plante}\columbia
\author{A.~C.~C.~Ribeiro}\coimbra
\author{R.~Santorelli}\columbia\zurich
\author{J.~M.~F.~dos Santos}\coimbra
\author{G.~Sartorelli}\bologna
\author{M.~Schumann}\zurich
\author{M.~Selvi}\bologna
\author{P.~Shagin}\houston
\author{H.~Simgen}\heidelberg
\author{A.~Teymourian}\losangeles
\author{D.~Thers}\subatech
\author{O.~Vitells}\weizmann
\author{H.~Wang}\losangeles
\author{M.~Weber}\heidelberg
\author{C.~Weinheimer}\munster

\collaboration{The XENON100 Collaboration}\noaffiliation
\begin{abstract}
The XENON100 experiment has completed a dark matter search
with 100.9 live days of data, taken from January to June 2010. Events
with energies between 8.4 and 44.6 keV$_{\n{nr}}$ in a fiducial volume
containing 48~kg of liquid xenon have been analyzed. A total of three
events have been found in the predefined signal region, compatible
with the background prediction of $(1.8\pm0.6)$ events. Based on this
analysis we present limits on the WIMP-nucleon cross section for
inelastic dark matter. With the present data we are able to rule out
the explanation for the observed DAMA/LIBRA modulation as being due to
inelastic dark matter scattering off iodine, at a 90\% confidence
level.
\end{abstract}



\maketitle

The interaction rate of dark matter particles from the Galactic halo is
expected to have an annual modulation, induced by Earth's motion
around the Sun~\cite{drukier1986}. Such a modulation has in fact been
observed in the DAMA/LIBRA
experiment~\cite{bernabei2008c,bernabei2010}. It is however difficult
to interpret this result as a signal from dark matter Weakly
Interacting Massive Particles (WIMPs), given the null results from
other direct dark matter searches~\cite{savage2009b}. In order to
overcome these tensions, inelastic dark matter (iDM) has been
proposed~\cite{smith2001,tuckersmith2004} as a modification of the
elastic WIMP model. iDM assumes that WIMPs scatter off baryonic matter
by simultaneously transitioning to an excited state at an energy
$\delta$ above the ground state ($\chi N\rightarrow\chi^*N$), while
elastic scattering is forbidden or highly suppressed. This introduces
a minimum velocity for WIMPs to scatter in a detector with a deposited
energy $E_{nr}$~\cite{chang2008}
$$
\beta_{min}=\sqrt{\frac{1}{2M_NE_{nr}}}\left(\frac{M_NE_{nr}}{\mu}+\delta\right),
$$ where $M_N$ is the mass of the target nucleus, $\mu$ is the reduced
mass of the WIMP/target nucleus system and $\delta$ is the energy
difference between the ground and excited state of the WIMP. In
particular, WIMPs with velocities lower than $\sqrt{2\delta/\mu}$ will
not be able to scatter at all since the kinetic energy is not
sufficient to allow the transition to the excited state. Therefore,
the available fraction of WIMPs that can interact will be larger for
more massive target nuclei, like iodine or xenon.

In contrast to elastic WIMP scattering, where an exponential recoil
energy spectrum is expected~\cite{Lewin:1995rx}, the velocity
threshold of the inelastic scattering process leads to a spectrum in
which the low energy component is suppressed and which peaks at
non-zero recoil energies. The recoil energy at which the rate is
maximal depends on $\delta$ and M$_\chi$. The differential event rate
is given by
$$
\frac{dR}{dE_{\n{nr}}}=N_TM_NA^2F^2\frac{\rho_\chi\sigma_N}{2M_\chi\mu^2}\int_{\beta_{\n{min}}}^\infty{\frac{f(v)}{v}dv},
$$ where $N_T$ is the total number of nuclei in the target, $A$ is the
atomic number of the target nucleus, $F$ is the nuclear form factor,
$\sigma_N$ is the WIMP-nucleon cross section and $\rho_\chi$ and
$M_\chi$ are the WIMP density and mass, respectively. $f(v)$ is the
halo velocity distribution function.  Another consequence of this
minimum velocity is the higher sensitivity of the recoil spectrum to
the tail of the WIMP velocity distribution, which enhances the annual
modulation effect for inelastic over elastic WIMP scattering.

The XENON100 experiment~\cite{Aprile:2010um} has recently reported
results from a 100.9~live days dark matter
search~\cite{spinindependent} in an energy interval between 8.4 and
44.6 keV$_{\n{nr}}$ (keV nuclear recoil equivalent). The same data are
used here to constrain the iDM model.  Three events fall in the
pre-defined WIMP search region for dark matter interactions, which is
compatible with the background expectation of $(1.8\pm0.6)$ events, as
described in~\cite{spinindependent}. 

To extract the DAMA/LIBRA allowed region in iDM parameter space, the
procedure described in~\cite{savage2009b} has been followed, using an
energy independent quenching factor of 0.08 for iodine and not
considering ion channeling. The DAMA/LIBRA modulation amplitudes for
different energies have been taken from~\cite{savage2009b}, where they
are extracted from figure 9 of~\cite{bernabei2008c}. Data have been
grouped in 17 bins, of which the last one corresponds to the energy
interval between 10 and 20~keVee. Different values of $\sigma_n$,
$\delta$ and $M_\chi$ have been selected and for each of them the
expected modulation amplitude in the DAMA/LIBRA experiment has been
computed. The DAMA/LIBRA allowed region is then defined as those
parameters for which $\chi^2$($M_\chi$, $\delta$)$<24.77$ for some
value of $\sigma_n$, where 24.77 corresponds to the value that is
excluded at 90\% confidence level for a $\chi^2$ distribution with 17
degrees of freedom.

\begin{figure}[ht]
\begin{center}\includegraphics[width = 1.0\columnwidth]{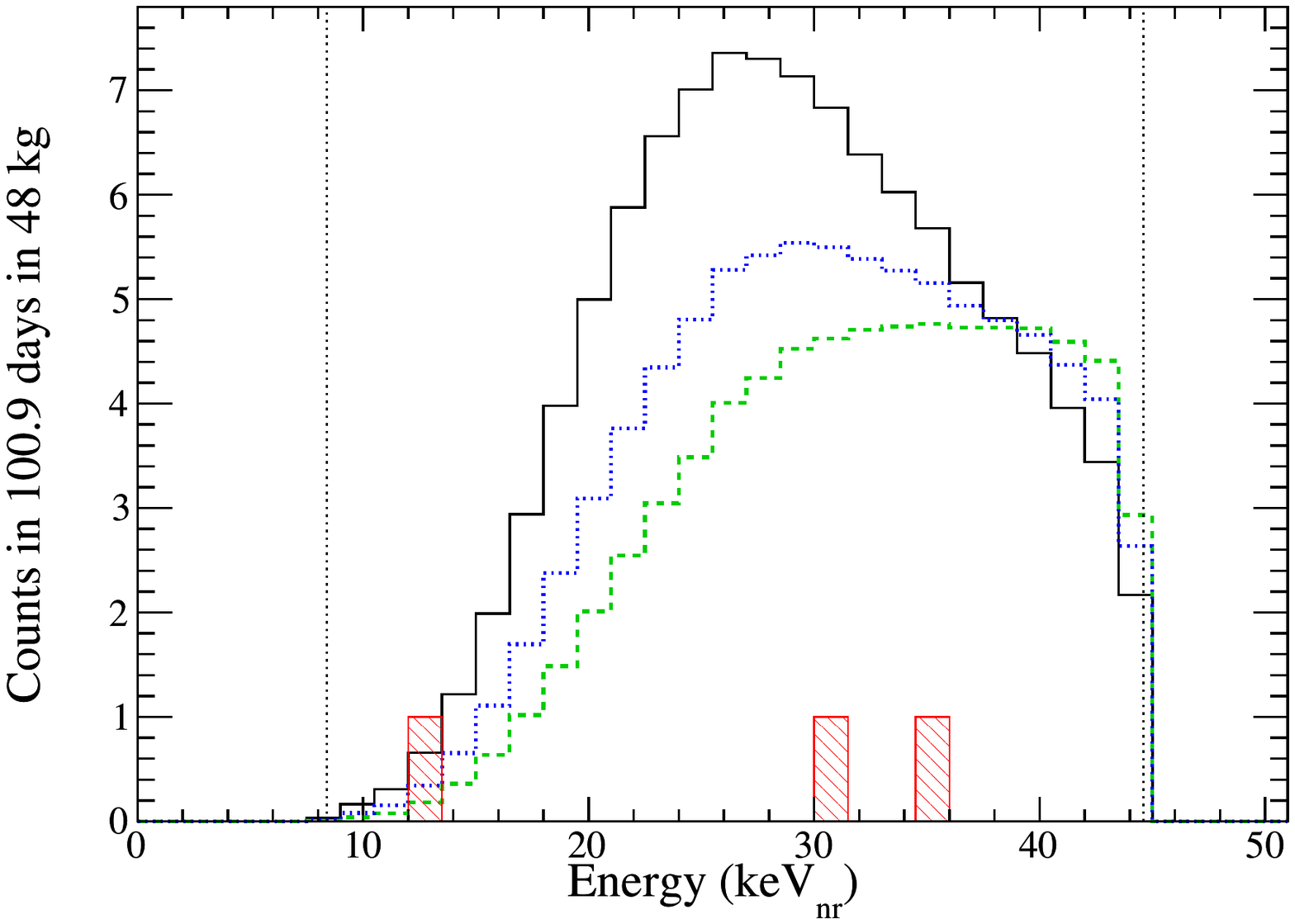}
\vspace{-20pt}
\caption{Expected iDM nuclear recoil spectrum in XENON100 for 100.9
  live days measured between January and June for a WIMP with
  $M_\chi=50~\n{GeV}$, $\delta=110\,\n{keV}$ (black, solid);
  $M_\chi=55~\n{GeV}$, $\delta=115\,\n{keV}$ (blue, dotted), and
  $M_\chi=60~\n{GeV}$, ~$\delta=120\,\n{keV}$ (green, dashed) and a
  $\sigma$ corresponding to the lower 90\% confidence limit of the
  DAMA/LIBRA signal. The XENON100 observed spectrum is shown in
  red. Vertical dotted lines show the analysis
  energy interval.}\label{fig:XENON100spectrum}
\end{center}\end{figure}

Following this procedure it is possible to compute for every point in
the allowed region the lowest cross section which is compatible with
DAMA/LIBRA at 90\% confidence level. The resulting cross section can
be used to predict a scatter rate in XENON100 and this can be compared
with the actual rate measured in XENON100. As an example to illustrate
the difference between the predictions from the DAMA/LIBRA data,
figure~\ref{fig:XENON100spectrum} shows the expected spectrum in
XENON100, taking into account exposure and data quality acceptance,
and the 90\% confidence level cross section from DAMA/LIBRA, for
different choices of M$_\chi$ and $\delta$ in the allowed region. The
WIMP velocity has been averaged over the data taking period to account
for annual modulation effects.

\begin{figure}[ht]
\begin{center}\includegraphics[width = 1.0\columnwidth]{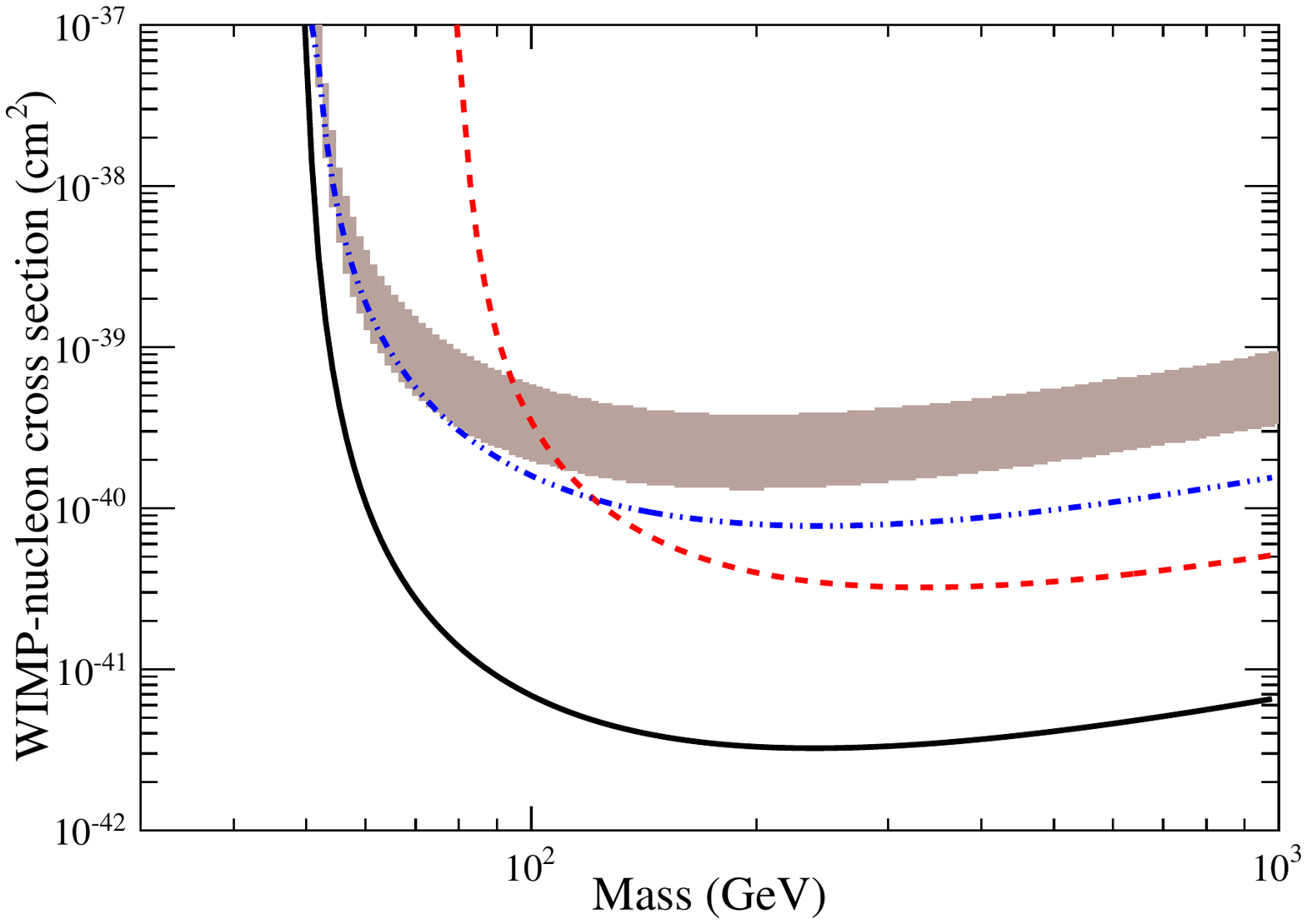}
\caption{DAMA/LIBRA 90\% confidence level signal region for
  $\delta=120~\n{keV}$ (gray region).  Superimposed are the 90\%
  confidence level exclusion curves for XENON100 (black, solid),
  CDMS~\cite{Ahmed:2009zw} (red, dashed) and
  ZEPLIN-III~\cite{akimov2010} (blue, dash-dotted). The whole
  DAMA/LIBRA WIMP region is excluded by
  XENON100.}\label{fig:DAMA120keV}
\end{center}\end{figure}

With this data a limit on $\sigma_N$ can be extracted for every pair
of $M_\chi$ and $\delta$ values using both the Feldman-Cousins
method~\cite{feldman} and the optimum gap method~\cite{yellin2002}. We
assume a Maxwellian WIMP velocity distribution with characteristic
velocity $v_0=220\1{km/s}$ and escape velocity
$v_{\n{esc}}=544\1{km/s}$, a local WIMP density of $0.3\1{GeV/cm^3}$,
Earth's velocity $v_{\earth}=29.8\1{km/s}$~\cite{savage2009b} and Helm
form factors~\cite{helm1956}.  Figure~\ref{fig:DAMA120keV} shows the
extracted limit for $\delta = 120$~keV using the Feldman-Cousins
method. The 90\% confidence region explaining the DAMA/LIBRA
modulation is also shown. It is excluded by the new XENON100 limit at
90\% confidence level.

\begin{figure}[ht]
\vspace{-10pt}
\begin{center}\includegraphics[width=1\columnwidth]{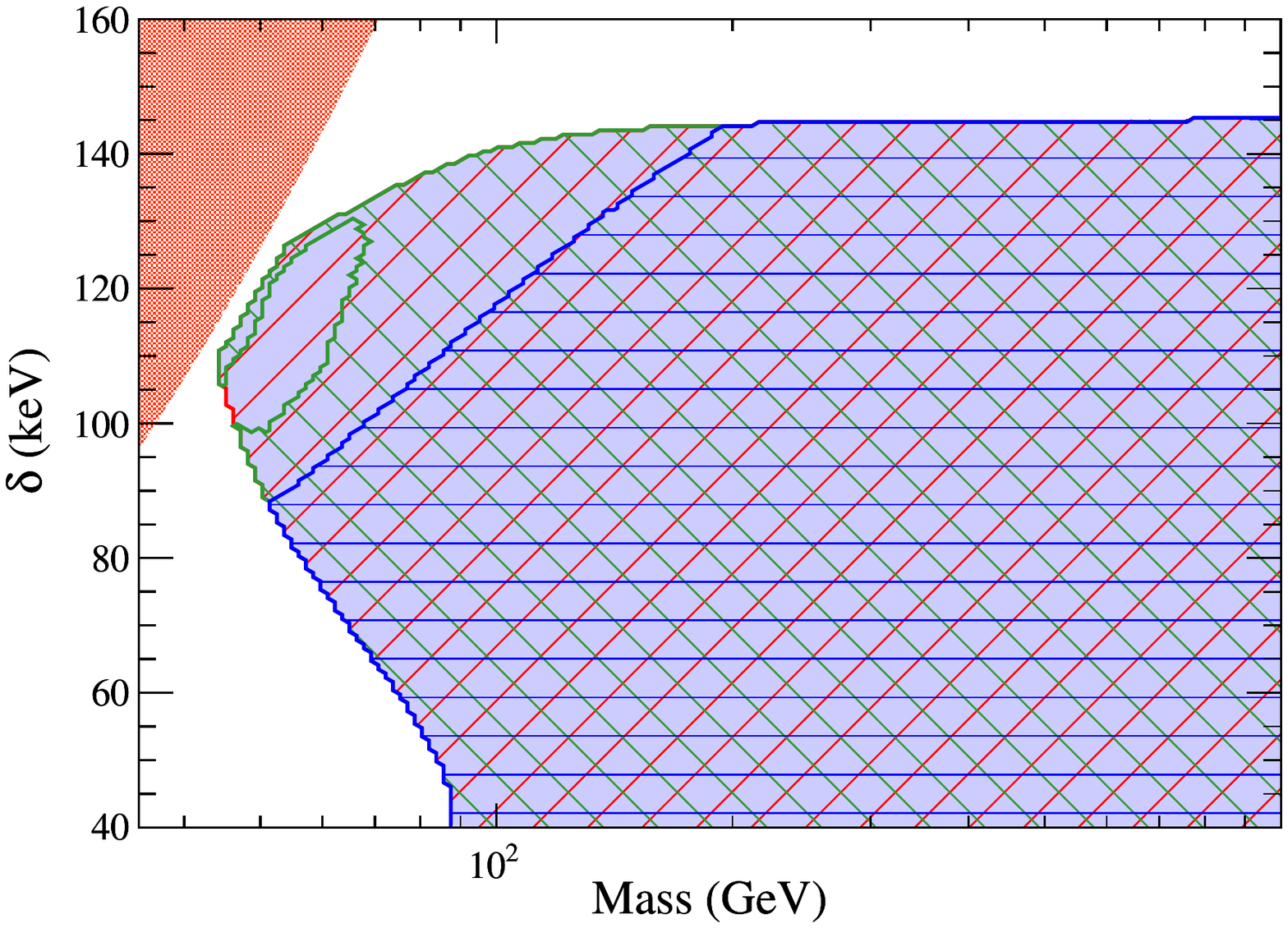}
\vspace{-20pt}\caption{Parameter space to explain the DAMA/LIBRA
  annual modulation with iDM (light blue area), and parameter space
  excluded by the CDMS-II~\cite{Ahmed:2009zw} experiment (blue
  horizontal lines), the ZEPLIN-III~\cite{akimov2010} experiment
  (green descending lines).  XENON100 (red ascending lines) excludes
  the whole allowed DAMA/LIBRA region. The orange region corresponds
  to the parameter space which is not accessible to any xenon
  experiment. $v_0=220\,\n{km/s}$ and
  $v_{esc}=544\,\n{km/s}$ have been assumed.}\label{fig:deltamass}
\end{center}\end{figure}

The systematic application of this procedure to the DAMA/LIBRA data
for all points in the $\delta$-$M_\chi$ space results in the gray area
in figure~\ref{fig:deltamass}, which shows the allowed parameter
space. To compare this result with other experiments, for each allowed
point in the $\delta$-$M_\chi$ space the lowest cross section in the
90\% signal region for the DAMA/LIBRA data is compared with the 90\%
confidence level limit cross section predicted by the other
experiment. In case the value from DAMA/LIBRA is higher than for the
experiment compared, that point in the parameter space is excluded.

Previous constraints from CDMS~\cite{ahmed2010b, Ahmed:2009zw},
CRESST~\cite{schmidthoberg2009} and EDELWEISS-II~\cite{edelweiss}
involve target nuclei with different masses than iodine, and thus
sample a different region of the WIMP velocity distribution. Thanks to
the similar mass of xenon and iodine, constraints inferred from liquid
xenon experiments are robust with respect to uncertainties in the
astrophysical parameters. This has already been shown by
ZEPLIN-III~\cite{akimov2010} and XENON10~\cite{Angle:2009xb}. These
data, however, left a small fraction of the spectrum available for
iDM, due to the limited exposure. With the XENON100 data the whole
DAMA/LIBRA parameter space is incompatible with the iDM explanation at
$90\percent$ confidence level. This result is independent of the
statistical method used to analyze the data.

Due to the cutoff at low energies associated with the iDM
interactions, the results can strongly depend on the chosen
astrophysical parameters. To ensure the robustness of the present
result, the calculations have been repeated for $v_{esc} =
500\,\n{km/s}$ and $v_{esc} = 600\,\n{km/s}$. The conclusion remains
unchanged. A source of systematic uncertainty often discussed in
liquid Xenon experiments is the conversion between measured light and
nuclear recoil energy, the so called
$\mathcal{L}_{\text{eff}}$~\cite{Plante:2011}. For this study,
however, this effect is very small due to the larger energies of
inelastic interactions compared to elastic ones.

An alternative explanation for the DAMA/LIBRA annual modulation based
on iDM WIMPs scattering off the Tl impurities in the NaI(Tl) crystals
has recently been proposed~\cite{Chang:2010pr}. Due to the small mass
of Xe compared with that of Tl, it is not possible to further
constrain the allowed parameter space than already done by the results
of the CRESST~\cite{schmidthoberg2009} experiment.

% ACKNOWLEDGEMENTS

We gratefully acknowledge support from NSF, DOE, SNF, Volkswagen
Foundation, FCT, R\'egion des Pays de la Loire, STCSM, DFG, Minerva
Gesellschaft and GIF.  We are grateful to LNGS for hosting and
supporting XENON.


\begin{thebibliography}{21}
\expandafter\ifx\csname natexlab\endcsname\relax\def\natexlab#1{#1}\fi
\expandafter\ifx\csname bibnamefont\endcsname\relax
  \def\bibnamefont#1{#1}\fi
\expandafter\ifx\csname bibfnamefont\endcsname\relax
  \def\bibfnamefont#1{#1}\fi
\expandafter\ifx\csname citenamefont\endcsname\relax
  \def\citenamefont#1{#1}\fi
\expandafter\ifx\csname url\endcsname\relax
  \def\url#1{\texttt{#1}}\fi
\expandafter\ifx\csname urlprefix\endcsname\relax\def\urlprefix{URL }\fi
\providecommand{\bibinfo}[2]{#2}
\providecommand{\eprint}[2][]{\url{#2}}

\bibitem[{\citenamefont{Drukier et~al.}(1986)\citenamefont{Drukier, Freese, and
  Spergel}}]{drukier1986}
\bibinfo{author}{\bibfnamefont{A.~K.} \bibnamefont{Drukier}},
  \bibinfo{author}{\bibfnamefont{K.}~\bibnamefont{Freese}}, \bibnamefont{and}
  \bibinfo{author}{\bibfnamefont{D.~N.} \bibnamefont{Spergel}},
  \bibinfo{journal}{Phys. Rev.} \textbf{\bibinfo{volume}{D33}},
  \bibinfo{pages}{3495} (\bibinfo{year}{1986}).

\bibitem[{\citenamefont{Bernabei et~al.}(2008)}]{bernabei2008c}
\bibinfo{author}{\bibfnamefont{R.}~\bibnamefont{Bernabei}} \bibnamefont{et~al.}
  (\bibinfo{collaboration}{DAMA}), \bibinfo{journal}{Eur. Phys. J.}
  \textbf{\bibinfo{volume}{C56}}, \bibinfo{pages}{333} (\bibinfo{year}{2008}).

\bibitem[{\citenamefont{Bernabei et~al.}(2010)}]{bernabei2010}
\bibinfo{author}{\bibfnamefont{R.}~\bibnamefont{Bernabei}}
  \bibnamefont{et~al.}, \bibinfo{journal}{Eur. Phys. J.}
  \textbf{\bibinfo{volume}{C67}}, \bibinfo{pages}{39} (\bibinfo{year}{2010}).

\bibitem[{\citenamefont{Savage et~al.}(2009)\citenamefont{Savage, Gelmini,
  Gondolo, and Freese}}]{savage2009b}
\bibinfo{author}{\bibfnamefont{C.}~\bibnamefont{Savage}},
  \bibinfo{author}{\bibfnamefont{G.}~\bibnamefont{Gelmini}},
  \bibinfo{author}{\bibfnamefont{P.}~\bibnamefont{Gondolo}}, \bibnamefont{and}
  \bibinfo{author}{\bibfnamefont{K.}~\bibnamefont{Freese}},
  \bibinfo{journal}{JCAP} \textbf{\bibinfo{volume}{0904}}, \bibinfo{pages}{010}
  (\bibinfo{year}{2009}).

\bibitem[{\citenamefont{Tucker-Smith and Weiner}(2001)}]{smith2001}
\bibinfo{author}{\bibfnamefont{D.}~\bibnamefont{Tucker-Smith}}
  \bibnamefont{and} \bibinfo{author}{\bibfnamefont{N.}~\bibnamefont{Weiner}},
  \bibinfo{journal}{Phys. Rev.} \textbf{\bibinfo{volume}{D64}},
  \bibinfo{pages}{043502} (\bibinfo{year}{2001}).

\bibitem[{\citenamefont{Tucker-Smith and Weiner}(2005)}]{tuckersmith2004}
\bibinfo{author}{\bibfnamefont{D.}~\bibnamefont{Tucker-Smith}}
  \bibnamefont{and} \bibinfo{author}{\bibfnamefont{N.}~\bibnamefont{Weiner}},
  \bibinfo{journal}{Phys. Rev.} \textbf{\bibinfo{volume}{D72}},
  \bibinfo{pages}{063509} (\bibinfo{year}{2005}).

\bibitem[{\citenamefont{Chang et~al.}(2009)\citenamefont{Chang, Kribs,
  Tucker-Smith, and Weiner}}]{chang2008}
\bibinfo{author}{\bibfnamefont{S.}~\bibnamefont{Chang}},
  \bibinfo{author}{\bibfnamefont{G.~D.} \bibnamefont{Kribs}},
  \bibinfo{author}{\bibfnamefont{D.}~\bibnamefont{Tucker-Smith}},
  \bibnamefont{and} \bibinfo{author}{\bibfnamefont{N.}~\bibnamefont{Weiner}},
  \bibinfo{journal}{Phys. Rev.} \textbf{\bibinfo{volume}{D79}},
  \bibinfo{pages}{043513} (\bibinfo{year}{2009}).

\bibitem[{\citenamefont{Lewin and Smith}(1996)}]{Lewin:1995rx}
\bibinfo{author}{\bibfnamefont{J.~D.} \bibnamefont{Lewin}} \bibnamefont{and}
  \bibinfo{author}{\bibfnamefont{P.~F.} \bibnamefont{Smith}},
  \bibinfo{journal}{Astropart. Phys.} \textbf{\bibinfo{volume}{6}},
  \bibinfo{pages}{87} (\bibinfo{year}{1996}).

\bibitem[{\citenamefont{Aprile et~al.}(2010)}]{Aprile:2010um}
\bibinfo{author}{\bibfnamefont{E.}~\bibnamefont{Aprile}} \bibnamefont{et~al.}
  (\bibinfo{collaboration}{XENON100}), \bibinfo{journal}{Phys. Rev. Lett.}
  \textbf{\bibinfo{volume}{105}}, \bibinfo{pages}{131302}
  (\bibinfo{year}{2010}).

\bibitem[{\citenamefont{Aprile et~al.}(2011)}]{spinindependent}
\bibinfo{author}{\bibfnamefont{E.}~\bibnamefont{Aprile}} \bibnamefont{et~al.}
  (\bibinfo{collaboration}{XENON100}) (\bibinfo{year}{2011}),
  \eprint{1104.2549}.

\bibitem[{\citenamefont{Feldman and Cousins}(1998)}]{feldman}
\bibinfo{author}{\bibfnamefont{G.~J.} \bibnamefont{Feldman}} \bibnamefont{and}
  \bibinfo{author}{\bibfnamefont{R.~D.} \bibnamefont{Cousins}},
  \bibinfo{journal}{Phys. Rev.} \textbf{\bibinfo{volume}{D57}},
  \bibinfo{pages}{3873} (\bibinfo{year}{1998}).

\bibitem[{\citenamefont{Yellin}(2002)}]{yellin2002}
\bibinfo{author}{\bibfnamefont{S.}~\bibnamefont{Yellin}},
  \bibinfo{journal}{Phys. Rev.} \textbf{\bibinfo{volume}{D66}},
  \bibinfo{pages}{032005} (\bibinfo{year}{2002}).

\bibitem[{\citenamefont{Helm}(1956)}]{helm1956}
\bibinfo{author}{\bibfnamefont{R.~H.} \bibnamefont{Helm}},
  \bibinfo{journal}{Phys. Rev.} \textbf{\bibinfo{volume}{104}},
  \bibinfo{pages}{1466} (\bibinfo{year}{1956}).

\bibitem[{\citenamefont{Ahmed et~al.}(2010{\natexlab{b}})}]{Ahmed:2009zw}
\bibinfo{author}{\bibfnamefont{Z.}~\bibnamefont{Ahmed}} \bibnamefont{et~al.},
  (\bibinfo{collaboration}{CDMS}),
  \bibinfo{journal}{Science} \textbf{\bibinfo{volume}{327}},
  \bibinfo{pages}{1619} (\bibinfo{year}{2010}{\natexlab{b}}).

\bibitem[{\citenamefont{Akimov et~al.}(2010)}]{akimov2010}
\bibinfo{author}{\bibfnamefont{D.~Y.} \bibnamefont{Akimov}}
  \bibnamefont{et~al.} (\bibinfo{collaboration}{ZEPLIN-III}),
  \bibinfo{journal}{Phys. Lett.} \textbf{\bibinfo{volume}{B692}},
  \bibinfo{pages}{180} (\bibinfo{year}{2010}).

\bibitem[{\citenamefont{Ahmed et~al.}(2010{\natexlab{a}})}]{ahmed2010b}
\bibinfo{author}{\bibfnamefont{Z.}~\bibnamefont{Ahmed}} \bibnamefont{et~al.}
  (\bibinfo{collaboration}{CDMS}),
  \bibinfo{journal}{Phys. Rev.} \textbf{\bibinfo{volume}{D83}},
  \bibinfo{pages}{112002} (\bibinfo{year}{2011}).

\bibitem[{\citenamefont{Schmidt-Hoberg and Winkler}(2009)}]{schmidthoberg2009}
\bibinfo{author}{\bibfnamefont{K.}~\bibnamefont{Schmidt-Hoberg}}
  \bibnamefont{and} \bibinfo{author}{\bibfnamefont{M.~W.}
  \bibnamefont{Winkler}}, \bibinfo{journal}{JCAP}
  \textbf{\bibinfo{volume}{0909}}, \bibinfo{pages}{010} (\bibinfo{year}{2009}).

\bibitem[{\citenamefont{Armengaud et~al.}(2011)}]{edelweiss}
\bibinfo{author}{\bibfnamefont{E.}~\bibnamefont{Armengaud}}
  \bibnamefont{et~al.} (\bibinfo{collaboration}{EDELWEISS})
  (\bibinfo{year}{2011}), \eprint{1103.4070}.

\bibitem[{\citenamefont{Angle et~al.}(2009)}]{Angle:2009xb}
\bibinfo{author}{\bibfnamefont{J.}~\bibnamefont{Angle}} \bibnamefont{et~al.}
  (\bibinfo{collaboration}{XENON10}), \bibinfo{journal}{Phys. Rev.}
  \textbf{\bibinfo{volume}{D80}}, \bibinfo{pages}{115005}
  (\bibinfo{year}{2009}).

\bibitem[{\citenamefont{Plante et~al.}(2011)}]{Plante:2011}
\bibinfo{author}{\bibfnamefont{G.}~\bibnamefont{Plante}} \bibnamefont{et~al.}
  (\bibinfo{year}{2011}), \eprint{1104.0000}.

\bibitem[{\citenamefont{Chang et~al.}(2011)\citenamefont{Chang, Lang, and
  Weiner}}]{Chang:2010pr}
\bibinfo{author}{\bibfnamefont{S.}~\bibnamefont{Chang}},
  \bibinfo{author}{\bibfnamefont{R.~F.} \bibnamefont{Lang}}, \bibnamefont{and}
  \bibinfo{author}{\bibfnamefont{N.}~\bibnamefont{Weiner}},
  \bibinfo{journal}{Phys. Rev. Lett.} \textbf{\bibinfo{volume}{106}},
  \bibinfo{pages}{011301} (\bibinfo{year}{2011}).

\end{thebibliography}
\end{document}